\documentclass[10pt,a4paper]{article}
\usepackage[top=1in,left=1in,footskip=1in,marginparwidth=1in]{geometry}

\usepackage{amsmath}
\usepackage{graphicx}
\usepackage{hyperref}

\begin{document}

\begin{centering}
{\bf{\Large Simplicity within biological complexity}}
\\
\bigskip
Nata\v{s}a Pr\v{z}ulj\textsuperscript{1,2,3,$\ast$},
No\"{e}l Malod-Dognin\textsuperscript{1}
\\
\bigskip
{1} Barcelona Supercomputing Center, 08034 Barcelona, Spain
\\
{2} Department of Computer Science, University College London, WC1E6BT London, United Kingdom
\\
{3} ICREA, Pg. Llu\'is Companys 23, 08010 Barcelona, Spain
\\
\bigskip
* natasha@bsc.es

\end{centering}

\bigskip

\begin{abstract}Heterogeneous, interconnected, systems-level, molecular (multi-omic) data have become increasingly available and key in precision medicine. We need to utilize them to better stratify patients into risk groups, discover new biomarkers and targets, repurpose known and discover new drugs to personalize medical treatment. Existing methodologies are limited and a paradigm shift is needed to achieve quantitative and qualitative breakthroughs. In this perspective paper, we survey the literature and argue for the development of a comprehensive, general framework for embedding of multi-scale molecular network data that would enable their explainable exploitation in precision medicine in linear time.  Network embedding methods (also called graph representation learning) map nodes to points in low-dimensional space, so that proximity in the learned space reflects the network’s topology-function relationships. They have recently achieved unprecedented performance on hard problems of utilizing few omic data in various biomedical applications. However, research thus far has been limited to special variants of the problems and data, with the performance depending on the underlying topology-function network biology hypotheses, the biomedical applications and evaluation metrics. 
The availability of multi-omic data, modern graph embedding paradigms and compute power call for a creation and training of efficient, explainable and controllable models, having no potentially dangerous, unexpected behaviour, that make a qualitative breakthrough.  We propose to develop a general, comprehensive embedding framework for multi-omic network data, from models to efficient and scalable software implementation, and to apply it to biomedical informatics, focusing on precision medicine and personalized drug discovery. It will lead to a paradigm shift in computational and biomedical understanding of data and diseases that will open up ways to solving some of the major bottlenecks in precision medicine and other domains.
\end{abstract}

\section*{Motivation}\label{sec1}
This century is the century of biology \cite{glover2012,venter2014}, characterized by biotechnological advances that 
continue to amass various molecular and clinical data \cite{oughtred2021,barrett2012,moreno2022,papatheodorou2020,pistachio,kearnes2021,reaxis,gene2023,baron2023,uffelmann2021,tcga,aacr2017,
lappalainen2012,ukbiobank}. Algorithmic and computational developments follow this data revolution striving to model and compute on the amassed data to uncover new biomedical knowledge \cite{prvzulj2019}. Versatile computational methods, also of ever-growing complexity, are being proposed to capture these ever-growing data and are incurring enormous computational and energy costs. Despite these substantial scientific labor and energy costs, the fundamental biomedical breakthroughs still elude us. Hence, we question whether omics data science is currently lost in its complexity and argue that it is the \emph{simplicity} that is hidden within the complexity of the data that we should endeavor to uncover. Recall that it is \emph{simplicity} (and not complexity) that has as an epistemological value long been taken as self-evident in science, a well-known philosophical principle illustrating it being Occam's razor \cite{chaitin2004,lichacz2021}.

However, the currently mainstream, black-box artificial intelligence / machine learning (AI / ML) models are not only excessively complex and incurring substantial energy costs, but also have unexpected behaviours, leading not only to some significant dangers, but also to surprising scientific discoveries, which we cannot control or account for \cite{higham2023,10.1007/978-3-031-44207-0_44}. Hence, we should not rely on such models to draw conclusions from human health-related data, but seek to uncover simple and fundamental principles underlying the molecular organization of life that determine the higher order behavior \cite{dhar2010,glazier2022,eftimie2022} and that we can easily and cheaply exploit. Currently there are only hints of such organizing principles and further developments are needed towards pioneering paradigm shifts to improve our understanding of the fundamental principles of biology, ultimately leading to the theory of molecular organization of life. We survey the state-of-the-art and propose modelling paradigms to facilitate the understanding of the simple underlying principles of multi-omics data organization leading to their controlled and sustainable exploitation and much needed scientific breakthroughs.


\section*{Current State-of-the-Art}\label{sec2}
\subsection*{Multi-omics data}\label{sec2.1}
\begin{figure}
\centering
\includegraphics[width=8.2cm]{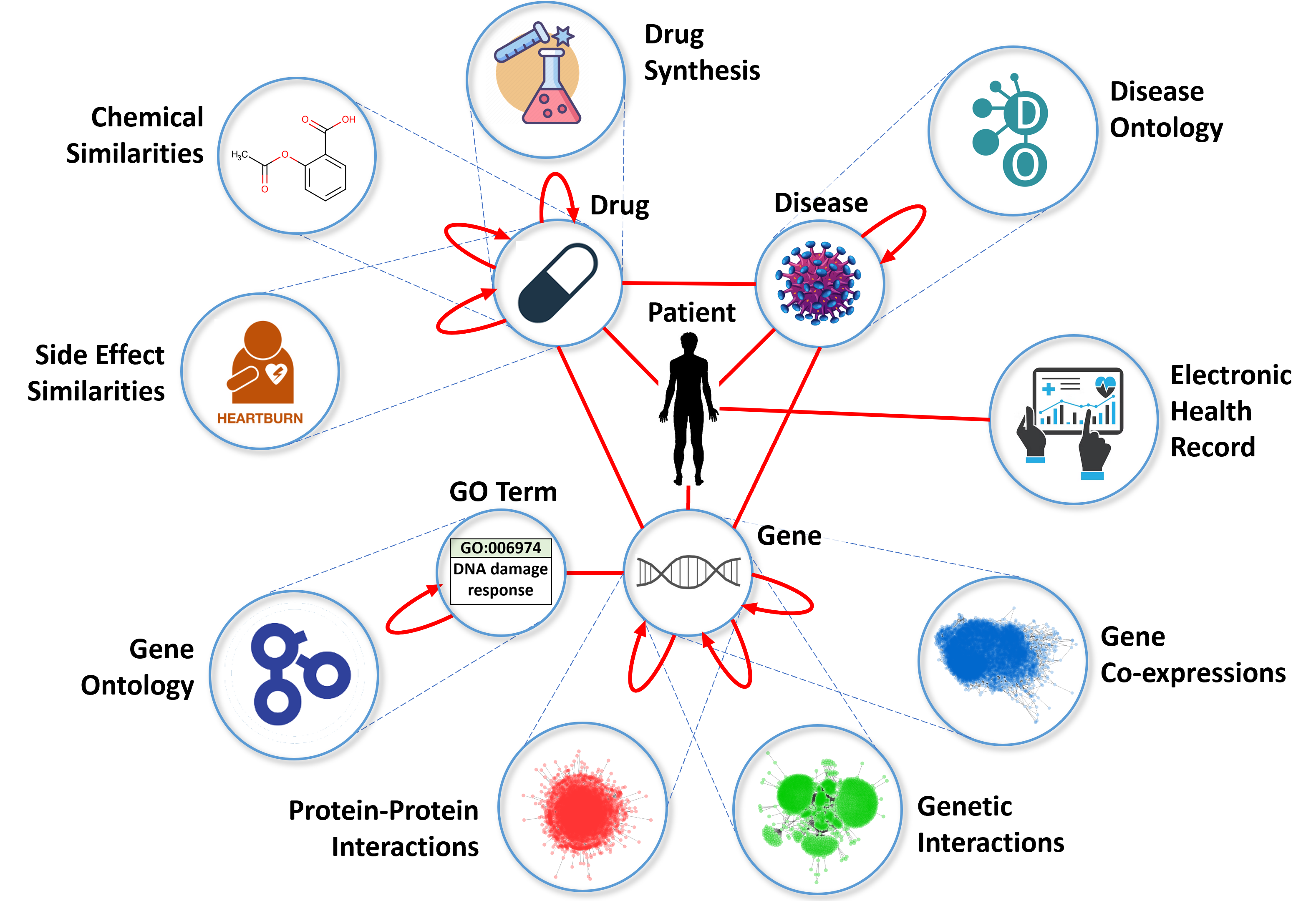}
\caption{Illustration of connectedness of biomedical network data.}\label{fig1}
\end{figure}

Modern biotechnologies have uncovered a complex system of heterogeneous interacting molecular entities, including genes, proteins and metabolites; they interact within the cells, bodies and with the external environment to maintain life and function. E.g., next-generation sequencing (NGS) techniques \cite{naturengs} have unlocked genome sequencing at mass scale and made it affordable. An integrated encyclopedia of DNA elements in the human genome (ENCODE) \cite{encode,encode2012,kellis2014,encodechan} has systematically mapped DNA regions of transcription, transcription-factor association, chromatin structure and histone modification, assigning biochemical functions for 80\% of the genome outside of the well-studied protein-coding regions (genes) \cite{encode2012,kellis2014}. Yeast two-hybrid assays \cite{ito2000,uetz2000,giot2003,li2004,stelzl2005,simonis2009,aimc2011} and affinity purification with mass spectrometry \cite{gavin2006,krogan2006} are widely used high-throughput methods for identifying physical interactions (bindings) between proteins, protein-protein interactions (PPIs). Synthetic genetic array (SGA) analysis automates the combinatorial construction of defined mutants and enables the quantitative analysis of genetic interactions (GIs) that exist between pairs of genes which when mutated together produce a phenotype that is not expected from the phenotypes produced by mutations of each of the two genes individually \cite{costanzo2010,baryshnikova2010,costanzo2016}. Other experimental technologies, such as microarrays \cite{quackenbush2001,dahlquist2002}, RNA-sequencing technologies \cite{marioni2008,mortazavi2008,wang2009,hirschhorn2005,duerr2006,macosko2015}, various single-cell omics \cite{macosko2015,klein2015,stegle2015,linnarsson2016} etc., have enabled construction of other omics layers in a cell, e.g. the genome, epigenome \cite{encode,encode2012,kellis2014,encodechan}, transcriptome \cite{quackenbush2001,dahlquist2002,marioni2008,mortazavi2008,wang2009,hirschhorn2005,duerr2006,macosko2015}, proteome  \cite{ito2000,uetz2000,giot2003,li2004,stelzl2005,simonis2009,aimc2011,gavin2006,krogan2006}, metabolome \cite{kegg}, exposome (data on nutrition, toxic molecules and radiation) \cite{kato2011,kumar2012,lambin2012,niedzwiecki2019}, microbiome \cite{nihhmp}, allergome \cite{allergome}, foodome \cite{foodome}, data on all drugs, their chemical similarities and bindings to protein targets (drug-target interactions, DTIs) \cite{drugbank,chembl,pubchem}, chemical reactions \cite{pistachio,kearnes2021,reaxis}, pesticides \cite{chembl,npic,ppdb}, herbicides \cite{chembl}, protein sequence \cite{uniprot,wang2018}, structure \cite{burley2023} and engineering data \cite{wang2018peb}, functional annotations of genes (Gene Ontology, GO) \cite{gene2023} and diseases (Disease Ontology, DO) \cite{baron2023}, genome-wide association studies \cite{uffelmann2021}, mutational data from The Cancer Genome Atlas (TCGA) that also contains epigenomic, transcriptomic, proteomic, and clinical data for 32 cancers (a landmark dataset for multi-omics methods development) \cite{tcga}, Pan-Cancer Atlas and AACR Project GENIE (a largest public, clinico-genomic cancer dataset) \cite{aacr2017}, including genomic structural variation data \cite{lappalainen2012}, versatile phenomic data from electronic health records \cite{ukbiobank} etc. These data are multi-scale, multi-omics layers describing cells and health status (illustrated in Fig. \ref{fig1}), each of which is a result of a type of biotechnological measurement that produced it, each with its technological limitations and biases, and therefore measuring different aspects the functioning of our cells, tissues and bodies \cite{prvzulj2019}. Substantial scientific efforts have been focusing on how to analyse these data to learn more about biology and curing diseases.

\subsection*{Biological network analytics}\label{sec2.2}
Networks (graphs) are universal descriptors of systems of interacting elements in biomedicine. Networks contain nodes, representing biological entities (e.g., genes, diseases, drugs), and edges, representing pairwise relationships between the entities (e.g., physical, functional, chemical) \cite{prvzulj2019}. Widely used molecular networks include those capturing PPIs \cite{oughtred2021}, GIs \cite{oughtred2021}, co-expression of genes (COEX) in tissues \cite{barrett2012,moreno2022,papatheodorou2020}, or in single cells \cite{barrett2012,moreno2022,papatheodorou2020}, in health and disease. Analysing the wiring (also called topology, or structure) of networks by graph theoretic approaches to extract biological information contained in the wiring, mostly from a single type of a molecular network data, has been a topic of many studies \cite{prvzulj2019}. More recently, significant efforts on developing data fusion algorithms for collectively mining these interconnected multi-omic data have been made \cite{prvzulj2019,gligorijevic2016,malod2019,zambrana2021,xenos2023}. However, there is an increasing realization that ever-growing multi-omic data, followed by models and algorithms of ever-growing complexity are hitting the limits due to computational intractability (e.g., NP-hardness) and over-parametrization (e.g., see section ``\nameref{sec2.2.2}'' below), having large carbon footprints. This is calling for alternative conceptual, modelling and computational approaches, with those based on embedding of the data points into lower-dimensional spaces, that would allow for their easier computational analyses, being expected to have ground breaking impacts in transforming the study of complex biomedical data \cite{li2022,nelson2019}.

In particular, graph theoretic methods applied on these data have fuelled biomedical discoveries, from uncovering biomarkers and relationships between diseases \cite{guo2019,le2016,menche2015,sumathipala2019} to repurposing of drugs \cite{morselli2021,cheng2019a,cheng2019b}, with algorithmic innovations including graphlets (higher order graph substructures) \cite{prvzulj2004,prvzulj2007,milenkovic2008,windels2019,windels2022}, applications of random walks \cite{chen2020,wong2020,yang2018}, kernels \cite{geng2020} and network propagation \cite{veselkov2019}, having been used for capturing structural information from networks. Extracting these predetermined features from a network and feeding the feature vectors into AI / ML models is a common approach, despite handcrafting optimally predictive features across diverse types of omics networks and biomedical applications being challenging \cite{zhang2020}. Hence, network embeddings \cite{nelson2019}, also called graph representation learning \cite{li2022}, have emerged as leading, cornerstone approaches, viewed as a bridge between network topology and classical ML, since ML is mostly able to process items in vector space. However, their development is challenging, because biomedical networks are very heterogeneous (Fig. \ref{fig1}), noisy, incomplete, complex, have no node ordering or reference points, and usually contain biomedical text and other domain knowledge, which makes embedding tasks more complicated than in other application domains. E.g., classic deep-learning (DL) neural network (NN) methods cannot handle such diverse structural properties and complex interactions, because they are designed for fixed-size grids (such as matrices of pixels in images and tabular datasets), or optimized for text and sequences. Hence, there is an increasing need for a deeper combination of network topology and ML \cite{su2020} into new modelling paradigms for biomedical systems that would parallel the successes of DL, which when applied on images and sequences have revolutionized image analysis and natural language processing (NLP).

\subsection*{Network embedding in computational biology}\label{sec2.2}
Network embedding methods represent the network nodes (or larger graph substructures) as points in low-dimensional space, leveraging the topology of the network, so that the points representing nearby or similarly wired nodes are put close in the space. Then, the vector representations of these points in space (corresponding to network nodes), a.k.a. embeddings, or embedding vectors, are used as input in traditional ML methods for downstream analyses, to link the network structure to molecular phenotypes, biological functions, or disease states. There is a wide range of network embedding methods in computational biomedicine, too many to describe here, that could roughly be classified by: 

\subsubsection*{I. The type of algorithms for generating network embeddings.}\label{sec2.2.1} This is a large research field, which, despite significant efforts, is still lacking due to the computational intractability (e.g., NP-hardness) of the underlying problems on graphs. There are many diverse algorithms that can be summarized into 7 major categories: (1) graph theoretic based \cite{prvzulj2019} (e.g., node degree, betweenness, graphlet counts, clustering); (2) random walks, or network diffusion based \cite{chen2020,wong2020,yang2018} (e.g., Diffusion State Distance \cite{cao2013}, GraphWave \cite{donnat2018}) ; (3) topological data analysis (TDA) based (e.g., persistent homology \cite{edelsbrunner2008,hofer2017,carriere2020,hofer2020,hofer2019,bubenik2015,edelsbrunner2022} and Mapper \cite{nicolau2011,singh2007,wang2019,de2015,madhobi2019,bodnar2021}); (4) manifold learning, or non-linear dimensionality reduction (e.g., Isomap \cite{tenebaum2000}, Laplacian eigenmap \cite{belkin2001} and t-SNE \cite{van2008}); (5) shallow neural network (NN) embeddings (e.g., unsupervised Skip-gram \cite{mikolov2013}, semi-supervised node2vec \cite{grover2016}); (6) supervised graph neural networks (GNNs) (e.g., see \cite{han2019,ingraham2019,xie2019,gilmer2017}); and (7) generative NN models (e.g., variational graph autoencoder \cite{kipf2016}). Another rough categorization, often used in ML and Knowledge Discovery \& Data Mining (KDD) communities \cite{goyal2018,cui2019,su2020} without an emphasis on a particular application domain (which can be a drawback, as all these methods are heuristic due to computational intractability of the underlying problems that they approximately optimize, so their performance is strongly dependent on particular data), is based on commonly used models into: (a) factorization based; (b) random walk based, and (c) deep-learning based.

\subsubsection*{II. The applications of embedding algorithms.}\label{sec2.2.2} There are many diverse applications, including predicting: protein structures, chemical compounds, PPIs, drug-drug interactions (synergistic or antagonistic), DTIs, new disease-related genes (e.g., new biomarkers, drug targets), drugs to repurpose for different diseases, disease co-morbidities etc. \cite{li2022,su2020}. Hence, biomedical applications are more diverse and complex than those usually considered by application-agnostic ML\&KDD communities, that mostly focus on node classification, link prediction and community detection in general \cite{goyal2018,cui2019}. Even when an application of a mainstream ML method in biology seems simple, the complexity of the resulting model is more than mind-boggling: e.g., transformers (a DL model designed to process sequential input data, primarily used in NLP and computer vision) were recently used to predict atomic level protein structure from protein sequence by training a family of large language models (LLM) that utilize DL, with 15 billion parameters, on 138 million known protein sequences \cite{lin2023}.

In the same vein, LLMs have been used to embed genes and cells \cite{yang2022,cui2023,chen2023}. For instance, genePT \cite{chen2023} embeds genes based on their text summaries from NCBI's Gene database and embeds single cells from single-cell RNA-seq expression data by using Open AI's GPT 3.5, a LLM model having 175B parameters. Also, DeID-GPT \cite{liu2023} de-identifies medical text (electronic health records) by using OpenAI's GPT4, a LLM model having 1.5T parameters.
Not only these billion/trillion parameter models are computationally costly to train and use, incurring substantial carbon footprint \cite{faiz2023}, but also several concerns arose about their usage for real-world and biomedical applications.
First, recent studies showed that the performances of LLM-based methods are often on par with those of simpler, dedicated approaches \cite{liu2023e,kedzierska2023,boiarsky2023}. For instance, Bioarsky \emph{et al.} \cite{boiarsky2023} showed that for annotating single sells from single-cell RNA-seq data, a simple L1 logistic regression can perform favourably against LLM-based approaches, such as scBERT\cite{yang2022} and scGPT \cite{cui2023}.
Another ongoing controversy of LLM-based approaches is their tendency to produce ``hallucinations,'' outputs that look coherent, but are factually incorrect, or nonsensical \cite{huang2023}.
These hallucinations raise concerns over the reliability of LLMs in real-world scenarios, and thus, limit their practical deployment \cite{huang2023}.

\begin{figure}[!t]
\centering
\includegraphics[width=4cm]{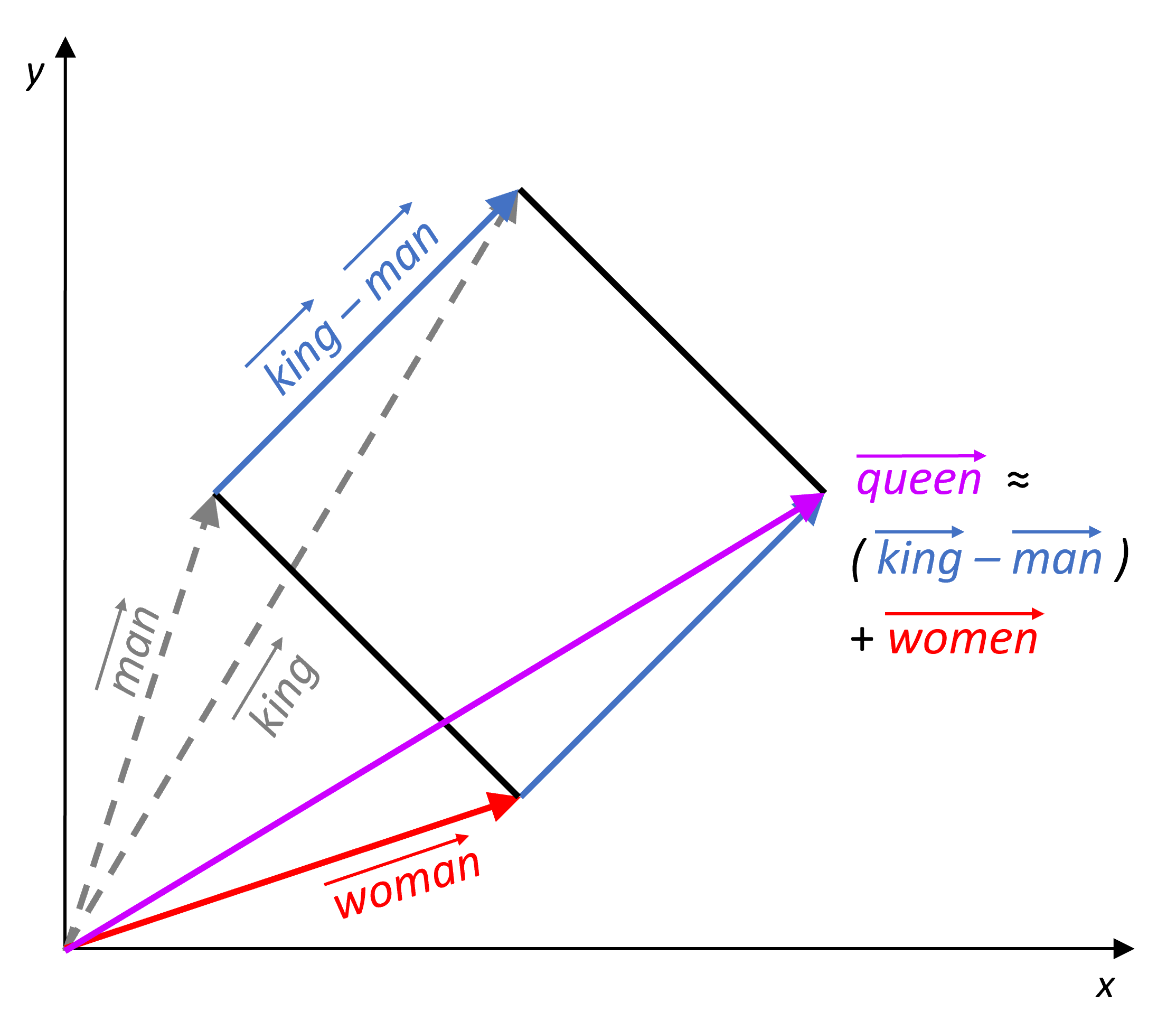}
\caption{Illustration of the linear analogy ``queen is to women what king is to man''.}\label{fig2}
\end{figure}

In addition to the above-mentioned limitations, LLMs also exhibit unexpected and surprising behaviour. For instance, Meta’s 15-billion-parameter model \cite{lin2023} (a competitor to RoseTTAFold \cite{baek2021} and Google’s AlphaFold \cite{jumper2021}) is trained to predict the next amino acid in sequence, but surprisingly it can also predict the corresponding 3D protein structure.
Similar unexpected behaviours leading to surprising discoveries happened in the field of NLP with utilization of NN-based embedding methods. For instance, recent successes in that field (document classification, named entity recognition, information retrieval) are attributed to the novel way of embedding words of text in a low-dimensional space, so that semantically similar words (e.g., synonyms, capital cities) are encoded by vector representations with low cosine similarity, e.g., by using Word2vec \cite{mikolov2013b} with the well-known Skip-Gram architecture, a 1-layer NN with a softmax function, seeking to maximize the dot product,  $\vec{w} \cdot \vec{c}$, of vectors $\vec{w}$  and $\vec{c}$, corresponding to word pairs, $(w,c)$,  in low $d$-dimensional embedding space that co-occur in the lexical corpus. As this architecture is computationally intractable for a large lexical corpus (e.g., the size of the widely used Google News corpus is billions of words), it has been replaced by the Skip-Gram with negative sampling (SGNS) \cite{mikolov2013}. Word vectors learned by the Skip-Gram model can meaningfully be combined by using just simple vector addition  \cite{mikolov2013}, enabling defining the vector representation of a phrase (sentence) as the average of the vector representations of its constituent words; this carries to paragraphs and entire texts \cite{quoc2014}. Surprisingly, the embedding space of words exhibits a linear algebraic structure: words represented as vectors in a low-dimensional space (word embeddings) can be used to extract new knowledge (analogies) directly by linear operations on the word vectors ($\vec{v}$) in the space, the famous example being that the vector calculation of $\overrightarrow{king}-\overrightarrow{man}+\overrightarrow{woman}$ is closer to $\overrightarrow{queen}$ than any other word vector (Fig. \ref{fig2}). However, since the first observations of such linear analogy properties of word embeddings \cite{mikolov2013c}, there have been controversies about the applicability of linear operations to solve real questions beyond simple examples \cite{rogers2017}. A key highlighted issue in these earlier approaches is the so-called ``offset problem:'' for instance, solving the above analogy by using the linear operation $\overrightarrow{queen}\approx \overrightarrow{king}-\overrightarrow{man}+\overrightarrow{women}$ supposes that the embedding vectors of these four words form a parallelogram (Fig. \ref{fig2}) and that any four words that are in the same relationship should also form a parallelogram, ignoring the fact that more important words (that appear more frequently in the text) tend to have longer embedding vectors than infrequent words. To overcome this limitation, machine learning was applied to uncover a suitable linear operation solving a particular analogy question beyond the simple parallelogram \cite{lim20211}.

\subsection*{Towards explainable, predictable and controllable AI models}
Other AI technologies also often have unexpected behaviours learned from the vast amounts of data that they analyse, so the AI community has begun to call for steps back from potentially dangerous, ever-larger, unpredictable, black-box models with emergent capabilities \cite{aaai2023,fli2023}. Clearly, new paradigms are needed to overcome these obstacles, to go beyond the controversies, unexpected behaviours leading to surprising discoveries and to provide understanding of the principles / laws underlying the organization of the data, which will make all observations and models explainable, predictable and controllable, hence enabling answering of all questions commonly asked from the data as simple applications of these laws. Elucidating the underlying laws and constructing explainable, controllable and environmentally sustainable models is the ambition to strive for. 

Inspired by the successes of word embedding methods in NLP, which revolutionized the field by embedding words into a low-dimensional space that exhibits linear algebraic structure (also having been applied to protein sequences, outlined in section ``\nameref{sec2.2.2}'' above) and enables solving word similarities and analogies, efficient document classification, retrieval etc., we hypothesize that it is also possible to embed all multi-omic data into a system of linearly connected, low-dimensional subspaces, each exhibiting linear algebraic structure. 
This could be viewed as taking projections of the entire high-dimensional space in which the biomedical entities reside, currently of unknown geometry and properties, into well-controlled and well-defined linear subspaces, which are linearly linked and traversed.
Such a system will enable computationally efficient answering of all commonly asked biomedical and precision medicine questions by applying linear operations on the embedding vectors of the biomedical entities from this system.

Such a system could be constructed by generalizing the observed linear algebraic structure of the embedding space of words in a lexical corpus from the field of NLP to biomedical multi-omics network embedding research, applied to address real problems in precision medicine.
As a proof of concept, we began with finding low-dimensional embeddings of various individual omics network layers into linear subspaces \cite{xenos2021,doria2023,doria2023b}.
Interestingly, we demonstrated that embedding the proteins of a PPI network by decomposing matrix representations of the network with Non-negative Matrix Tri-Factorization (NMTF, detailed below)  results in embedding spaces that are functionally organized. In particular, first we demonstrated that in the resulting embedding spaces, proteins with similar biological annotations (functions) are embedded close in the embedding space \cite{xenos2021,doria2023,doria2023b}. Second, we demonstrated that the embedding space itself is functionally organized, with axes of the space capturing new, data-driven, higher-order biological functions, which we uncovered by joint embedding of genes and annotations into the same space \cite{doria2023b}. 
Importantly, we initiated the exploitation of the linear organization of these embedding spaces to answer biological questions. For instance, we showed that by applying simple vector operations on the embedding vectors of proteins we can uncover new members of protein complexes and predict new cancer-related genes \cite{xenos2021}. As protein complexes in PPI networks are homophilic, meaning that adjacent proteins tend to perform the same biological function, direct application of NMTF to factorize the Positive Pointwise Mutual Information (PPMI, defined below) matrix of the PPI network based on random walks yielded this linearly exploitable subspace \cite{xenos2021}.  However, for heterophilic data, such as cancer driver genes (and equivalently proteins, as gene products), which have been known to exhibit similarity in the wiring patters in the PPI network even when being non-adjacent in the network (i.e., they are heterophilic), we had to introduce a new version of the PPMI matrix based on randomly walking between nodes with similar graphlet-based wiring patterns to obtain the linear behavior exploitable for predicting new cancer-related genes: we successfully predicted new cancer-related genes by performing linear operations on the embedding vectors of cancer driver genes in this new, graphlet-enabled embedding space \cite{xenos2021}.

In addition to embedding one or few omic network types, e.g. PPI networks described above, we should also be embedding integrated / fused multi-omics network layers, also into linear subspaces, and we should find linear transformations that link these linear subspaces, enabling answering of all commonly asked biomedical questions by linear operations on the low-dimensional embedding vectors of biomedical entities. That is, we should untangle, ``linearize'' and organize all multi-omics data into a low-dimensional, linear subspace system, enabling controllable and explainable answering of all common biomedical questions by simple linear operations on low-dimensional embedding vectors (hence, compute-space saving) of biomedical entities in linear time (hence, compute-time saving), enabling not only revolutionizing of biomedical data science, but also saving computational space and power, i.e., reducing the carbon footprint.
Encouraged by our recent preliminary feasibility studies that proposed a new, topology-constrained, matrix factorization-based embedding of the human PPI network into the low-dimensional embedding space exhibiting linear algebraic structure, which yielded new biomedical knowledge and identified new cancer-related genes and functions only by applying linear operations on the protein embedding vectors \cite{xenos2021,doria2023,doria2023b}, we expect that furthering such approaches will achieve quantitative and qualitative leaps both in modelling and algorithmic development and in precision medicine and other applications. 
We expect them to elucidate simple, linear organization of multi-omics data systems and enable effective utilization and exploitation of the multi-omic data in precision medicine.

To achieve that, we need new methodological advances towards a new representation and mining of the rich, multi-scale structure of multi-omic molecular organization within an integrated system, which will uncover new, emergent nature of biological organization where linear organization holds. Importantly, the advances should be applied to explore some of the most challenging problems in different realms (e.g., in precision medicine) where the proposed methods can utilize the wealth of available multi-omic data to help address one of the greatest problems in all areas of science, the nature of complex organization. 
To realize the ambition, the new methodologies should be based on the four scientific pillars described below.

\subsection*{Pillar I. Characterizing biological network structure}\label{sec:2.1}
Many computational problems on large networks are computationally intractable (e.g., subgraph isomorphism problem \cite{cook1971}, which underlies large network comparisons and makes them compute intensive). Also, because nature is variable and the biological data are noisy, traditional graph theoretic algorithms are of little use for network biology, and more flexible, intentionally approximate (heuristic) approaches are necessary. The area of developing such methods in biological and other domains is vast, too big to comprehensively survey here, some recent reviews including  \cite{liu2020,koutrouli2020,guo2022}. Hence, we summarize it and argue for extending only the state-of-the-art approaches relevant to demonstrating that the hypothesis of this manuscript holds.

Easily computable macroscopic statistical global properties of large networks have extensively been examined. The most widely used global network properties are the degree distribution, clustering coefficient, clustering spectra, network diameter and various forms of network centralities \cite{newman2003}. The degree of a node is the number of edges touching the node and the degree distribution is the distribution of degrees of all nodes in the network. Many large real-world networks have non-Poisson degree distributions with a power-law tail, termed scale-free \cite{barabasi1999}. However, networks with exactly the same degree distributions can have vastly different structure affecting their function \cite{prvzulj2004,li2005}. The same holds for other global network properties \cite{prvzulj2004}. Furthermore, global network properties of largely incomplete molecular networks do not tell us much about the true structure of these networks; instead, they describe the network structure produced by the sampling biotechnologies used to obtain these networks \cite{han2005,stumpf2005}. Thus, global statistics on such incomplete data may be substantially biased, or even misleading with respect to the (currently unknown) full network. Conversely, certain neighborhoods of these networks are well-studied (usually the regions of a network relevant for human disease). Since we have detailed knowledge of certain local areas of biological networks, but data outside these well studied areas is currently incomplete, local statistics are likely to be more valid and meaningful. Furthermore, biological experiments for detecting molecular and other biochemical networks are of local nature. Thus, many have focused on developing tools based on local network structure. The state-of-the-art such methods are based on graphlets, introduced in 2004 \cite{prvzulj2004} and defined as small, connected, non-isomorphic, induced subgraphs of large networks \cite{prvzulj2004}. Graphlets and the symmetry groups in them, called automorphism orbits (introduced in 2007 \cite{prvzulj2007}), have extensively been used to develop many new tools for analyzing structural properties of networks, some based on graphlet and orbit frequency distributions in a network \cite{prvzulj2007,milenkovic2008}, others including graphlet Laplacians, eigencentralities, spectral clustering \cite{windels2019,windels2022} etc. Graphlet statistics have also been used as kernels and feature vectors in various ML methods  \cite{shervashidze2009,vacic2010}, and in the message-passing framework of GNNs, yielding better results in downstream analysis tasks \cite{bouritsas2022}. They have been utilized and cited in around 21,000 research papers in various domains and in around 300 patents according to Google Scholar. Also, they were generalized to mine the multi-scale network organization: to hypergraphlets \cite{gaudelet2018,lugo2021} in hypergraphs, and simplets \cite{malod2019b} in abstract simplicial complexes. Due to their large applicability and the above-mentioned increasing need for a deeper combination of network topology and ML \cite{su2020} for modelling biomedical systems, we propose that extending and utilizing graphlet-based methodologies would achieve breakthroughs.

\subsection*{Pillar II. Omic network data fusion }\label{sec:2.2}
Omics network data fusion is another vast research area, too large to comprehensively survey here, recent surveys including \cite{agamah2022,reel2021,lee2022,flores2023,ritchie2015,gligorijevic2015,gligorijevic2016b,picard2021,vahabi2022}. Hence, after a brief general overview, we focus only on the sub-area relevant to the future research perspectives presented in this manuscript. 

A wide variety of techniques that leverage the information contained in the relationships between omics data types have been driven by different biomedical applications (e.g., patient subtyping, biomarker discovery, de novo drug discovery, drug repurposing, and personalizing of treatment) and various challenges specific to multi-omics data analyses (e.g., omic data heterogeneity and high-dimensionality). The multitude of approaches motivate several rough categorizations into broader methodological groups, e.g. into early, middle, and late integrative methods, alternatively called concatenation-based, transformation-based, or model-based integration, respectively \cite{ritchie2015,gligorijevic2015,gligorijevic2016b,picard2021}, because: early integration usually involves an initial concatenation of the features across the measured omics, followed by application of methods for analysing the resulting high-dimensional dataset; middle integration applies a transformation to represent a complex combination of the datasets before applying downstream analyses; a late integration method analyses each dataset separately and introduces a model or algorithm that combines the outputs of each individual analysis. Another characterization focuses on specific techniques, e.g. kernel learning, matrix factorization, graph/network representations, and deep learning. The categorizations are not mutually exclusive and a given method may be difficult to classify. Furthermore, use of ML to analyse multi-omic data face key, unique, data-driven challenges, roughly summarized as: (a) data heterogeneity, sparsity and outliers; (b) class imbalance and over-fitting; (c) much more features than the data (``curse of dimensionality''); (d) computational and storage cost; (e) choosing an algorithm that works best for a given biomedical problem is hard, since ML methods are approximate (heuristic) due to computational intractability (e.g., NP-hardness) of the underlying problems (e.g., non-linear optimization), so from the theory of computation we know that each method is guaranteed to fail on some particular examples, calling for a design of robust, application-specific methods. 

Matrix factorizations (MF) have largely demonstrated their usefulness in improving understanding of biological mechanisms from omic data, e.g.: Singular Value Decomposition (SVD) \cite{alter2000}, Principal Component Analysis (PCA) \cite{wall2003} with the sparse and probabilistic variants \cite{meng2016}, Independent Component Analysis (ICA) \cite{sompairac2019} and Nonnegative Matrix Factorizations (NMF) \cite{stein2018,yang2016,boccarelli2018,rappoport2018,esposito2019,esposito2021}. Each of these techniques is based on different constraints that characterize the final properties of the matrix factors, leading to different optimization problems and numerical algorithms being used. NMF has become a standard tool in the analysis of high-dimensional data, a comprehensive and up-to-date detail of the most important theoretical aspects, including geometric interpretation, non-negative rank, complexity, and uniqueness, having been provided \cite{gillis2020}. Importantly, a very broad array of models belongs to the ``factorized model'' family \cite{haeffele2015}. 

A particular variant for omics network fusion, Non-negative Matrix Tri-Factorization (NMTF) \cite{devarajan2008}, has been gaining in popularity. It was originally proposed for dimensionality reduction and co-clustering due to its connection with k-means clustering \cite{fei2008}. It decomposes an $n\times m$ data matrix, $A$ (e.g. the adjacency matrix of a biological network), representing the relations between $n$ and $m$ elements, into a product of three non-negative, low-dimensional matrices, $G_{n\times k}$, $S_{k\times k_1}$ and $P_{m\times k_1}$, as $A \approx GSP^T$, by solving the optimization problem: $min \left(||A - GSP^T||^2_F: G, S, P \geq 0\right)$. The low-dimensional matrix factors $G$ and $P$ are used to assign $n$ data points into $k\ll n$ clusters and $m$ data points into $k_1\ll m$ clusters, respectively (the clustering property of NMTF) \cite{ding2006}, and $S$ is a $k\times k_1$ compressed representation of $A$ linking clusters of $G$ and $P$, and also allowing for different numbers of clusters in $G$ and in $P$ ($k$ and $k_1$, respectively). $G$ and $P$ imply co-clustering. The reconstructed matrix $GSP^T$ is more complete then the initial data matrix $A$, featuring new entries, unobserved in the data, which emerge from the latent structure captured by the low-dimensional matrix factors, hence inferring new predictions (the matrix completion property of NMTF \cite{gligorijevic2016}). 

\begin{figure}
\centering
\includegraphics[width=8.2cm]{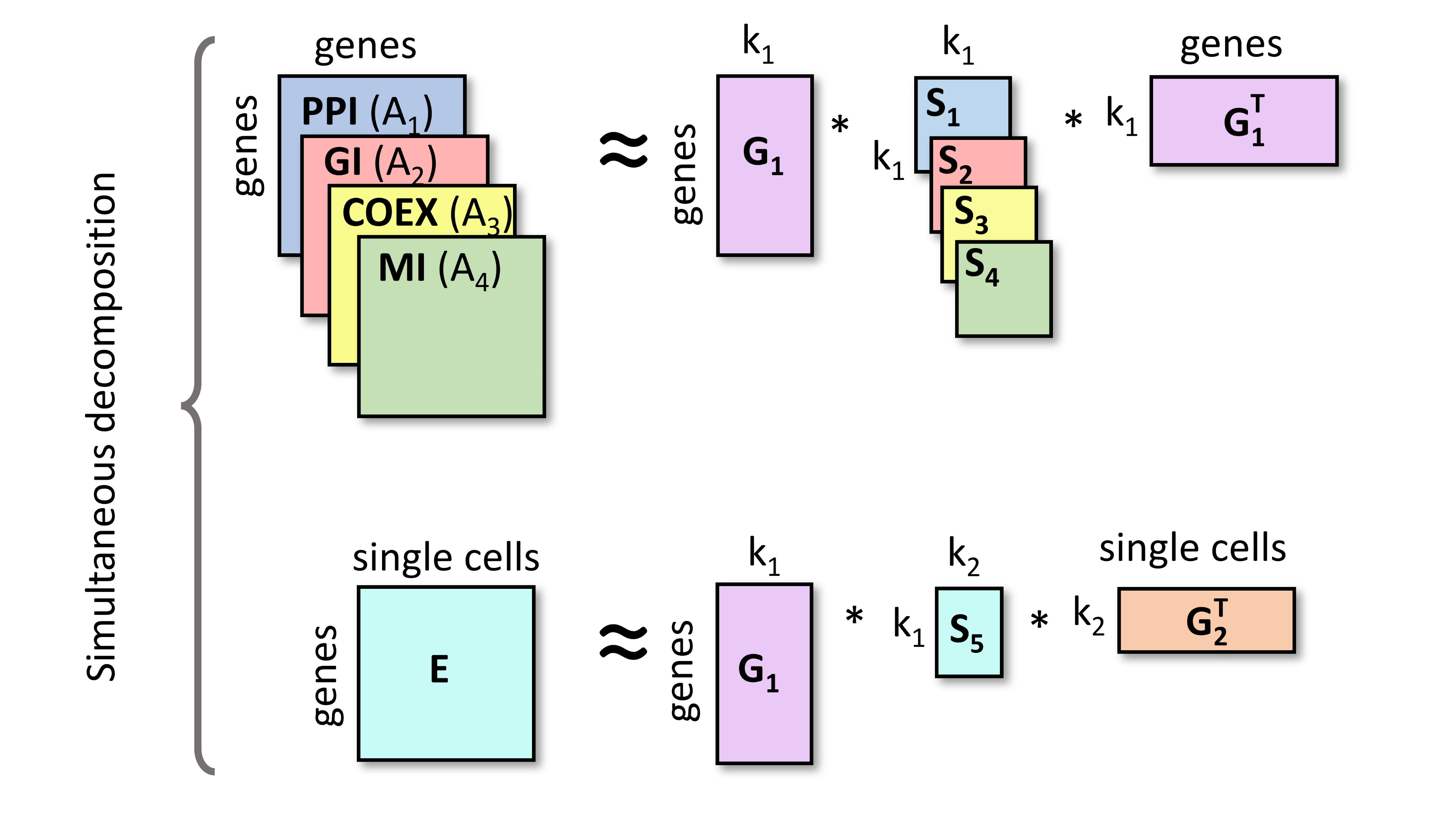} \\
\begin{equation*}
J_1 = \min_{G_1, G_2, G_3, H_{12}, H_{23} \geq 0} 	\left( \begin{array}{c}     || R_{12} - G_1H_{12}G_2^T||_F^2   \\
											+  || R_{23} - G_2H_{23}G_3^T||_F^2  \\
								 			+ tr(G_2^T L_2G_2) + tr(G_3^T L_3G_3) \\
								\end{array}	\right)	
\end{equation*}

\caption{Illustration of the NMTF model from \cite{zambrana2021}. The viral host interactions, VHIs (represented by matrix $R_{12}$), are simultaneously decomposed with the drug-target interactions, DTIs (represented by matrix $R_{23}$).
The matrix factor $G_2$ is shared across decompositions to allow learning from all input matrices. The first graph regularization penalty (illustrated by the green arrow) is added so that the human genes that interact in the molecular interaction network, MIN (represented by its Laplacian matrix, $L_2$), are assigned similar low dimensional embedding vectors in $G_2$. Similarly, the second graph regularization penalty (illustrated by the red arrow) is added so that the drugs that have similar chemical structures in the drug chemical similarity network, DCS  (represented by its Laplacian matrix, $L_3$), are assigned similar low dimensional embedding vectors in $G_3$. The hyper-parameters $k_1$, $k_2$ and $k_3$ indicate the reduced dimensions of the embedding spaces of the viral proteins, human proteins and drugs, respectively.
The lower dimensional matrix factors, $G_1$, $G_2$, $G_3$, $H_{12}$ and $H_{23}$ are obtained by solving the corresponding minimization problem, $J_1$.
}\label{fig3a}
\end{figure}

\begin{figure}
\centering
\includegraphics[width=8.2cm]{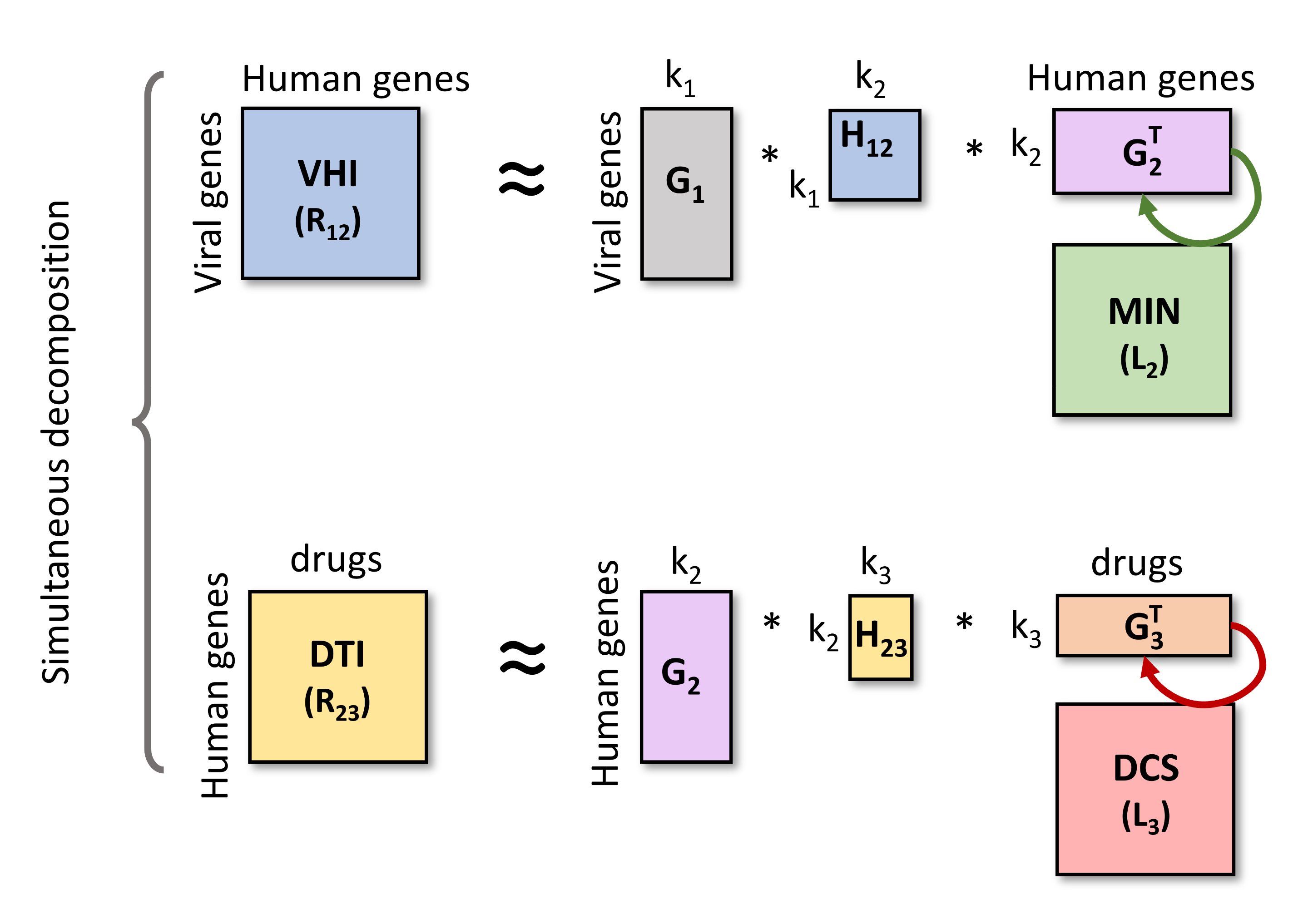} \\
~\\
\begin{equation*}
J_2 = \min_{G_1, G_2, S_1, S_2, S_3, S_4, S_5 \geq 0} \left( \begin{array}{c}	\sum_{i=1}^{4} || A_i - G_1S_{i}G_1^T||_F^2 \\
												 +  || E - G_1S_{5}G_2^T||_F^2\\
								\end{array} \right) 
\end{equation*}

\caption{Illustration of the NMTF model from \cite{mihajlovic2023m}. The single-cell gene expression matrix, $E$ (which can be thought of as capturing the phenotype), is simultaneously decomposed with four molecular interaction networks, PPI, GI, COEX and MI, represented by their adjacency matrices, $A_1$, $A_2$, $A_3$, and $A_4$, respectively (which can be thought of as capturing the genotype, as they describe all possible interactions). The matrix factor $G_1$ is shared across decompositions to allow learning from all input matrices. The hyper-parameters $k_1$ and $k_2$ indicate the reduced dimensions of the embedding spaces of human genes and single cells, respectively.
The lower dimensional matrix factors, $G_1$, $G_2$, $S_1$, $S_2$, $S_3$, $S_4$ and $S_5$ are obtained by solving the corresponding minimization problem, $J_2$.
}\label{fig3b}
\end{figure}

Importantly, NMTF allows for fusion of heterogeneous network data, by sharing matrix factors during decompositions.  For instance, as illustrated in Fig. \ref{fig3a}, to re-purpose known drugs for Covid-19, we recently factorized viral-host PPIs (VHI) simultaneously with drug-target interactions (DTIs) as $VHI \approx G_1H_{12}G_2^T$ and $DTI \approx G_2H_{23}G_3^T$, where $G_1$ matrix clusters $n$ viral proteins into $k_1$ clusters, $G_2$ clusters $m$ human proteins into $k_2$ clusters, $G_3$ clusters $q$ drugs into $k_3$ clusters, $G_2$ matrix is shared across the decomposition to ensure data fusion, and the reconstructed entities of the $G_2H_{23}G_3^T$ matrix resulting from data fusion identify new DTIs used to re-purpose drugs \cite{zambrana2021}. The only hyper-parameters to tune were low-dimensional $k_1$, $k_2$ and $k_3$, with a total of about 2.6M parameters (the number of entries in the low dimensional matrix factors $G_1$, $G_2$, $G_3$, $H_{12}$ and $H_{23}$).
As another example, we recently integrated single-cell gene expression data from patient-derived induced pluripotent stem cells together with four molecular interaction networks by using the NMTF model presented in Fig. \ref{fig3b} to uncover novel Parkinson's disease related genes and pathways \cite{mihajlovic2023m}. Again, and unlike the large language models (LLMs) that requiere billions/trillions of parameters to tune (e.g., \cite{lin2023,liu2023,cui2023}), our proposed NMTF model only requieres two hyper-parameters to tune, the low dimensionality parameters $k_1$ and $k_2$, and about 1.17M parameters (the number of entries in the low dimentional matrix factors $G_1$, $G_2$, $S_1$, $S_2$, $S_3$, $S_4$ and $S_5$).
These two examples illustrate that by careful modelling of the relationships between the data points, requiring deep understanding of the biomedical domain, with very few hyper-parameters to tune, NMTF can achieve results superior to those of highly parametrized ML methods \cite{xenos2021,doria2023,malod2019,zambrana2021,xenos2023,isokaanta2020,mihajlovic2023m}.

There are several other reasons for which NMTF has increasingly become a methodology of choice for multi-omics data fusion. First, non-negativity that applies to both bases and weights allows for meaningful interpretations. Second, modularity in the sense of the methodology easily enabling adding or removing omics data layers (illustrated by $A_1$ to $A_4$ and $E$ in Fig. \ref{fig3b}) is consistent with the natural human part-based learning processes \cite{esposito2021b}. Third, NMTF-based methods overcome the shortcomings of the competing ML data fusion methods (Bayesian network-based and Kernel-based integration \cite{gligorijevic2015b}): they model inter-type data relations (e.g. patients-to-drugs) and intra-type data relations (e.g., PPIs) simultaneously, require no data transformation (hence, incur no information loss), and incorporate both intra- and inter-type networks at their full systems-level size, into a single integrative model from which new medical knowledge is inferred \cite{gligorijevic2015,gligorijevic2016b}. Importantly, NMTF results in an integrated network in which: (i) some ``less important'' (or topologically inconsistent) links across the networks do not appear (e.g., links appearing only in one network); (ii) links with structural support in most of the networks appear in the integrated network with enhanced association scores; (iii) new links emerge as a consequence of the matrix completion property as predictions that need to be further examined. I.e., these methods give the highest weight to the information supported by evidence from multiple data, hence diminishing the effects of sparsity and noise in the data \cite{malod2023}.  Although NMTF is equivalent to the 2-factor, non-negative matrix-factorization (NMF) problem \cite{ding2006}, $min \left(||D - GF^T||^2_F : G, F \geq 0\right)$, which is NP-hard \cite{vavasis2010}, so we cannot expect to solve NMTF to optimality, availability of fast optimization methods \cite{vcopar2019} coupled with exploitation of sparsity of the omic data and utilization of parallel matrix manipulations on high-performance computing (HPC) facilities make these methods not only applicable, but also feasible for omics data fusion.

\subsection*{Pillar III. Omics network embedding methods (NLP-inspired)}\label{sec:2.3}
The area of network embedding methods is also too large to comprehensively survey here, with recent reviews including \cite{li2022,su2020,goyal2018,cui2019,xu2021}. They can roughly be divided into those based on traditional matrix factorizations (MF), neural networks (NNs), and random walks. The traditional MF-based methods come in many variants, e.g.: SVD \cite{alter2000}, Isomap \cite{tenebaum2000}, Locally Linear Embedding \cite{roweis2000}. Those based on NNs include: multilayer perceptrons \cite{tang2015}, autoencoders \cite{kipf2016}, generative adversarial networks \cite{wang2019b}, graph neural networks (GNN) \cite{zhou2020} including graph convolutional networks (GCN) \cite{kipf2017}, and geometric deep learning \cite{atz2021}. A limitation of these methods is in that they mainly focus on factorizing the first-order data matrix (adjacency matrix), despite limitations of representing networks with adjacency matrices and using as input into NNs, e.g. that many different adjacency matrices can represent the same graph connectivity, but do not produce the same result when input into a NN, i.e., they are not permutation invariant \cite{ying2018hierarchical}  (some works recently attempting to address it \cite{meltzer2019pinet,huang2022going,balan2022permutation}). Other limitations include: lack of robustness to noise and overfitting due to scarcity of labelled data, interpretability, and real networks usually being ``multiplex'' \cite{yang2020}, i.e. having incomplete multi-view representations from different relations (some works recently attempting to address it \cite{reel2021,lee2022,flores2023,wang2022d}). 

Random-walk-based network embedding methods are inspired by the field of NLP, where words are represented as vectors in a low, $d$-dimensional space (word embeddings) \cite{mikolov2013b} and new knowledge is extracted directly by linear operations on the vectors in the space. It has long been known that two words in a similar context have similar meanings (``distributional hypothesis'') \cite{harris1954}. Following this observation, Word2vec model was introduced \cite{mikolov2013b}, which generates continuous representations of words as vectors in $d$-dimensional space, so that words that appear frequently in the same context are placed close in the space (detailed above). However, this architecture is computationally intractable for a large lexical corpus (e.g., the size of the widely used Google News corpus is billions of words) and therefore has been replaced by the Skip-Gram with negative sampling (SGNS) \cite{mikolov2013} and more robust versions of it were proposed \cite{armandpour2019}. Importantly, this model enables semantic, context-based, word comparisons (``similarity task'') in the embedding space solely by computing the cosine similarity of their vector representations \cite{mikolov2013b}. E.g., Paris and Berlin, being capital cities, have similar vector representations and consequently, high cosine similarity. As noted above, word representations learned by the Skip-Gram model can also meaningfully be combined using vector addition \cite{mikolov2013}, enabling defining of the vector representation of a phrase (sentence) as the average of the vector representations of its constituent words \cite{le2014d}; more recent works (e.g., \cite{lin2017s}) train DL models to learn phrase representations that respect the word order in the sentence. These allow for analysing texts beyond the level of words, revealing semantic similarities between sentences, paragraphs, or entire documents. However, serious limitations exist about the applicability of linear operations to solve real questions beyond simple examples \cite{rogers2017}, a key issue being the above mentioned ``offset problem,'' among others. Analogously in networks, hub-nodes (of very high degree) occur on more random walks and hence, have longer embedding vectors than peripheral nodes. These limitations need to be addressed. 

Following the success of the Skip-Gram model, various attempts have been made to generalize it and apply it to networks, e.g., DeepWalk \cite{perozzi2014d}, LINE \cite{tang2015l}, PTE \cite{tang2015pte} and Node2vec \cite{grover2016}. These methods rely on random walks to generate sequences of nodes, the equivalent of the lexical corpus, on which the Skip-Gram architecture can be applied. Interestingly, it was shown that the SGNS is implicitly factorizing a word-context matrix \cite{levy2014n}, called Positive Pointwise Mutual Information (PPMI) matrix, defined as follows: for two words, $w$ and $c$,  $PPMI(w,c)=\max\left(0, \log{\frac{(w,c)|C|}{|w||c|}}\right)$, where $|C|$ is the size of the corpus, $(w,c)$ is the number of times the two words co-occur in the corpus, and $|w|$ and $|c|$ are the numbers of times the words $w$ and $c$ occur in the corpus. The cells of this matrix quantify how frequently two words of the lexical corpus co-occur in a sliding window compared to what would be expected if the occurrences of the words were independent. Subsequently, it was shown that the Skip-Gram-based network embeddings (in DeepWalk, LINE, PTE, and Node2vec) can be unified into the MF framework with closed forms \cite{qiu2018n}: they are implicitly factorizing a random-walk-based mutual information matrix, $M$ (a diffusion on the original network), which is equivalent to the above described PPMI matrix, as its cells quantify how frequently two nodes of the network co-occur in a random walk compared to what would be expected if the occurrences of the nodes were independent. Formally, each entry of $M$, $M_{ij}$, is the logarithm of the average probability that node $i$ randomly walks to node $j$ in a fixed number of steps. A closed formula to approximate $M$ was presented and used to generate the embedding space by applying SVD on matrix $M$ (NetMF method and its approximation algorithm for computing network embedding) \cite{qiu2018n}. This provided the theoretical connections between Skip-gram based network embedding algorithms and the theory of graph Laplacian \cite{qiu2018n}.

These methods embed network nodes in a low-dimensional space, so that nodes that are ``similar'' are close in space, where ``similarity'' means either belonging to the same network neighbourhood or ``community,'' or having similar network topology, e.g. being hub, or bridge nodes (structural / topological similarity). Then vector representations of nodes are used as input into ML models to predict protein functions, drug-disease associations, drug–drug interactions and PPIs (see \cite{su2020} for details). Hence, in both cases of words and networks, the Skip-Gram-based embeddings are approximating the exact factorization of the mutual information matrix. In NLP they use the NN-based embeddings rather than explicitly factorizing the PPMI matrix due to the size (factorizing a matrix of 1B$\times$ 1B words being computational intractable, as time complexity of SVD is $O(n^3))$. In contrast, the size of the human PPI network and thus its PPMI matrix is currently $\approx 19,000\times 19,000$, making its decompositions, either with SVD, or NMTF feasible (time complexity is $O(n^3)$). This motivates us to investigate whether the equivalent properties of word embeddings (molecular analogues to word-analogies, Fig. \ref{fig2}) also hold for molecular networks if we decompose their PPMI, or adjacency matrix representation, with an NMTF-based framework \cite{gligorijevic2016,malod2019,zambrana2021,xenos2023}. In omics networks, the best equivalent to word analogy are functional modules, e.g., protein complexes and pathways in PPI networks. However, biological function is not only shared between the proteins that physically interact \cite{sharan2007n} or participate in the same functional module \cite{chen2014i}, but also between proteins that have similar local wiring patterns regardless of their adjacency in the interaction network \cite{milenkovic2008}, best quantified by graphlet-based measures \cite{prvzulj2004,prvzulj2007,milenkovic2008,windels2019,windels2022}, e.g., graphlet degree vectors (GDVs) \cite{milenkovic2008}. Thus, an important challenge is to generate embeddings that locate close in space nodes with similar wiring patterns.

There are several other reasons for which we argue that NMTF-based framework is the methodology of choice to embed each of the omics network layers individually (e.g., PPI, GI, COEX) into an embedding space whose properties we can control, that exhibits linear algebraic structure, which enables answering of various bio-medical questions by linear operations on the embedding vectors of nodes, and which importantly enables fusion of all omics networks into a linearly-traversable system of embedding subspaces (detailed below). The 7 constraints to enforce control for the system to be free of potentially dangerous, unexpected behaviours are: 1) non-negativity of the coordinates of the embedding vectors (for easier interpretation), 2) bases (axes) of the spaces being orthogonal (to minimise entity dependencies in the embedding space), 3) sparsity (to capture the strongest signal), 4) enable data fusion, 5) enable ``walking'' (by linear transformations) between the embedding spaces as needed to answer biomedical questions, i.e., directly investigating different types of entities in the same, or in different linear subspaces (modelling different types of data), 6) explainability, capturing how the entities relate (i.e., being eXplainable AI, XAI), and 7) few parameters to tune (to simplify the model). NMTF-based methods \cite{ding2006} are the only that can easily satisfy all 7 constraints. They were successfully used in biology to analyse individual omics data types in isolation from each other, and also enable heterogeneous data fusion, including network topology as constraints, further motivating their utilization within new network embedding algorithms in  biomedical applications  \cite{gligorijevic2016,malod2019,vitali2018p}. Importantly, explainability (constraint 6 above) allows control of the system's behaviour to coincide with what is expected (by examining the linear operation that gave the answer), while reduction of unexpected behaviour comes mostly from orthogonality (constraint 2 above) and the property of NMTF that it boosts the signal that comes from multiple data types, rather than outliers, resulting in more robust predictions (constraint 4 above).

\subsection*{Pillar IV. Finding optimal dimensionality of the embedding spaces}\label{sec:2.4}
Another main challenge in graph embeddings is finding an optimal reduced dimension that is small enough to be efficient, but large enough to keep all of the necessary properties of the whole network. The choice of the dimension considerably influences the model’s performance \cite{lai2016g}, with the embeddings with very low dimensionality typically not being expressive enough to capture the richness of the data \cite{patel2017t}, while embeddings with a very large dimensionality suffering from over-fitting. Also, large dimensionalities tend to increase model complexity, slow down training speed, and add inferential latency, all of which are constraints that can potentially limit model applicability and deployment \cite{wu2016g}. As this is an unresolved scientific question, the embedding space dimensionality is considered a hyper-parameter of the model. In most existing methods, the dimensionality is selected by a grid search, by estimating the performance in downstream tasks, such as node classification, or link prediction. The choice of the optimal dimension is mainly evaluated for non-biological networks, with few studies evaluating it on biological networks \cite{yue2020g}. The approaches that estimate the intrinsic dimensionality of the data usually yield super low dimensionality (at most 20), while the data-driven methods yield up to 200/250 dimensions \cite{yin2018d}. Recently proposed ultra-low dimensions (20 or fewer) are shown to preserve network structure, but perform poorly in downstream classification tasks \cite{chanpuriya2020n}. The popular approaches using hyperbolic space to embed networks are not applicable for omics networks, because they assume that the number of squares or pentagons is higher than the number of triangles, which is not the case for our data \cite{almagro2022d,seshadhri2020i}. The vast majority of analyses that use Node2vec \cite{grover2016} and DeepWalk \cite{perozzi2014d} use the default dimensionalities of these methods, 128 or 256. Our initial, data-driven feasibility studies on PPI networks demonstrate that the lower the dimensionality, the less distinguishable the embedding vectors of genes, and as we increase the number of dimensions, the average distance between the embedding vectors increases and hence the space is more disentangled \cite{xenos2021,doria2023}, but after 250/300 dimensions the position of the embedding vectors in the embedding space does not change, suggesting this to be an upper limit for PPI networks \cite{xenos2021,doria2023}. 
Recently, the 2NN method \cite{facco2017estimating} for estimating the dimensionality of datapoints in high-dimensional space has been successfully extended for estimating the dimensionality of network embedding spaces \cite{grindrod2023estimating}. While this approach has yet to be tested on real biological networks, it may be the basis of novel measures for estimating the optimal dimensionality of the system of embedding subspaces used to integrate the multi-omics datasets in precision medicine.

\section*{Perspectives and Challenges}
We propose that multi-disciplinary efforts are needed to address the above challenges and develop an advanced, explainable and controllable network embedding methodology, free of potentially dangerous, unexpected behaviour. It will marry biomedical informatics with network science, linear algebra and data fusion, unlocking foremost emerging interdisciplinary fields, precision medicine and personalized drug discovery, enabling extraction of new medical knowledge from all data collectively. To achieve this, innovation is needed in the following main areas.

\subsection*{Innovation in methods for embedding multi-omic networks into a liner sub-space system}\label{sec:3.1}
We should aim at bridging the gap between multi-omic network data and their network embedding-based biomedical interpretability by generalizing and uniting NMTF-based network embedding and network-science methods to propose new algorithmic and biological paradigms and solve real problems in precision medicine and other domains. We should generalize sophisticated network embedding methods to encompass and model the multi-scale structure of molecular organization within a linearly-traversable system of embedding spaces exhibiting linear algebraic structure. This pre-trained system should fuse all publicly available multi-omics data types and serve as a basis to uncover new biomedical paradigms and knowledge. It will enable us to find solutions to these generalizations that are as good as possible (locally optimal), which can be achieved by using the best practices from non-linear optimization and the best HPC infrastructure, which given NP-hardness of these problems, is the best we can do. Encouraged by our recent feasibility studies yielding significant results for cancer \cite{xenos2021,doria2023,doria2023b},  we expect that such approaches will achieve quantitative and qualitative leaps in computational development and applications.

\subsection*{Innovation in data-science and biological paradigms}\label{sec:3.2}
The new methods will enable much needed paradigm shifts in data science and biology. In data science, by simultaneously capturing the adjacency and higher-order structural network information along with the semantic information (node labels) within the new state-of-the-art network embedding algorithms (see section ``\nameref{sec:3.1}'' above), it will be possible to describe a general methodology for uncovering homophilic representations of multi-labelled networks,  leading to functionally organized, linearly separable, embedding spaces suitable for applying linear operations. This would pave the road to generating a new generation of embedding methods that could drastically improve the accuracy of downstream analysis tasks for heterophilic data, which is a bottleneck of the current embedding methods. This, along with the proposed orthogonality of the embedding spaces (to minimize dependencies) \cite{xenos2021,doria2023,doria2023b,malod2023,gureghian2023multi},  would further enable a shift in the exploration of the embedding spaces from objects’ embedding vectors to the orthonormal axes of the space \cite{doria2023b}.  In the biological domain, it would enable the much called for displacement of the dominant, sequence alignment-based construction of GO \cite{yu2019d} and the species phylogenetic organization \cite{foley2023g}, resolving the controversial relationships and clarifying diversification (e.g., in the evolution of SARS-CoV-2 virus that caused the recent pandemic) by the proposed multi-omics data-driven approaches, guaranteed to provide better disentanglement of the functional and phylogenetic information.

\subsection*{Applications in precision medicine, drug discovery and other domains}\label{sec:3.3}
The proposed advances can be applied and validated on real multi-omic data. They will enable innovation in utilizing new AI frameworks, encompassing multi-omics ``big data,'' to enable continued growth in digital innovation applied to precision medicine. They will enable better subtyping of patients, identifying new biomarkers, drug-targets and drugs to repurpose, expending beyond oncology, particularly in neurologic diseases affecting the aging population: e.g. Alzheimer's disease (AD), Parkinson’s disease (PD) \cite{mihajlovic2023m}, and glioblastoma cancer (GBM). The proposed directions would enable AI-assisted personalized drug discovery \cite{mendez2020n,yasonik2020m,segler2018g,grechishnikova2021t,popova2018d} (drugs for particular patients/patient subgroups) simultaneously with the other precision medicine tasks, at the same time and cost, from analysis of the same multi-omics data, within the same linear system (see sections ``\nameref{sec:3.1}'' and ``\nameref{sec:3.2}'' above), unique thus far, further enabling better insights in medicine and drug discovery. Transformative potential of the proposed directions can also be envisioned in agriculture \cite{nicolopoulou2016c,blois2022f,mei2022p,pires2022c,orvsolic2021c} and protein engineering \cite{wang2018p,greenhalgh2021,freschlin2022}.


\section*{Competing interests}
No competing interest is declared.

\section*{Author contributions statement}
N.P. conceived the conceptual framework and the philosophy of the research perspectives presented in the manuscript. N.P. and N.M-D. wrote and reviewed the manuscript.

\section*{Acknowledgments}
This work is supported by the European Research Council (ERC) Consolidator Grant 770827, the Spanish State Research
Agency and the Ministry of Science and Innovation MCIN grant PID2022-141920NB-I00 / AEI /10.13039/501100011033/ FEDER, UE, and the Department of Research and Universities
of the Generalitat de Catalunya code 2021 SGR 01536.

\bibliographystyle{unsrt}
\bibliography{Paper}

\end{document}